\documentclass[11pt]{article}

\usepackage[a4paper,margin=1in]{geometry}
\usepackage{amsmath,amssymb,amsthm,mathtools}
\usepackage{hyperref}
\usepackage{graphicx}
\usepackage{enumitem}
\usepackage{float}
\usepackage{tikz}
\usetikzlibrary{arrows.meta,positioning,calc}

\newcommand{\ket}[1]{\lvert #1\rangle}
\newcommand{\bra}[1]{\langle #1\rvert}
\newcommand{\proj}[1]{\ket{#1}\!\bra{#1}}
\newcommand{\id}{\mathrm{id}}
\newcommand{\Tr}{\mathrm{Tr}}

\theoremstyle{definition}
\newtheorem{definition}{Definition}[section]
\newtheorem{remark}{Remark}[section]

\theoremstyle{plain}
\newtheorem{proposition}{Proposition}[section]
\newtheorem{lemma}{Lemma}[section]

\title{Bell-like States in Classical Optics:\\
A Process-Theoretic and Sheaf-Theoretic (Categorical) Clarification}
\author{Partha Ghose\thanks{Email: partha.ghose@gmail.com}\\
Tagore Centre for Natural Sciences and Philosophy,\\
Rabindra Tirtha, New Town, Kolkata 700156, India}
\date{}

\begin{document}
\maketitle

\begin{abstract}
Classical polarization optics is naturally described by a two-dimensional complex Hilbert space
(Jones vectors). Hence the monoidal tensor-product kinematics used to define bipartite
nonseparability is already available in a classical setting. When one adds intrinsic stochasticity
(statistical optical fields) and adopts an operational stance in which outcomes need not be
pre-assigned prior to detection, Bell--CHSH correlations of quantum strength can be obtained for
suitably prepared two-beam polarization states.  Beyond its conceptual role, the stochastic-optics platform provides a tunable, low-cost testbed for probing the robustness of Bell/CHSH and contextuality witnesses against realistic imperfections—noise, coarse measurement binning, and selective sampling.

An alternative preparation based on external conical refraction (ECR), in which intersecting conical-refraction rings mimic the intersecting emission cones of SPDC, is also outlined.

This paper gives a self-contained categorical
formulation of the situation. The experimental arrangement (Hadamard-like splitting,
CNOT-like coupling, and routing/conditioning that physically removes unwanted contributions) is
treated as a single morphism in an operational process theory (e.g.\ $\mathbf{CPM}(\mathbf{FHilb})$).
From this process description one functorially extracts an empirical model: a compatible family of
context-indexed probability distributions. The Abramsky--Brandenburger sheaf-theoretic criterion
then applies: noncontextuality is equivalent to the existence of a global section, and CHSH
violation is a precise failure-to-glue statement.  The two-level analysis separates kinematic
nonseparability from operational contextuality and clarifies why neither, by itself, entails nonlocal
causation. Conceptually, this shows that contextuality (failure of a global section) is not synonymous with
microscopic quantumness: it can arise in a classically implementable stochastic-optics regime.
\end{abstract}


\section{Introduction}

Hilbert spaces are not exclusive to quantum theory. Classical polarization optics is naturally
modelled by normalized Jones vectors, i.e.\ a two-dimensional complex Hilbert space for the
polarization degree of freedom. Tensor products then enter in two distinct but closely related ways:
(i) \emph{intrastate} (or ``intrasystem'') nonseparability within a single beam, e.g.\ polarization$\otimes$spatial mode,
which underlies much of the ``classical entanglement'' literature \cite{Kagalwala2013,QianEtAl2015,ForbesAielloNdagano2019,Spreeuw1998,GhoseMukherjee2013review};
and (ii) \emph{interstate} nonfactorisability between two distinct beams, e.g.\ polarization$\otimes$polarization,
which is the setting relevant for Bell/CHSH-type correlation tests \cite{GhoseSamal2001}.

Mathematically, the nonseparability at issue is governed by the Schmidt decomposition (equivalently, a singular-value decomposition), a theorem formulated independently of quantum mechanics (in fact, in 1907, before its advent) \cite{Schmidt1907} and applicable wherever a complex inner-product space and a tensor factorisation are used. Historically, however, the word ``entanglement'' entered physics discourse through Schr\"odinger\'s 1935 response to the EPR scenario \cite{EinsteinPodolskyRosen1935,Schrodinger1935}, which helped cement an association between nonseparability, quantumness, and (in the interstate case) nonlocality. It is therefore worth keeping in mind that intrasystem
entanglement is also routine in quantum mechanics (e.g.\ path$\otimes$spin in single-neutron interferometry) \cite{HasegawaErdosi2011}. In short: both classical optics and quantum theory allow intra- and inter-system nonseparability at the level of Hilbert-space kinematics; what differs is what these structures enable operationally.

A recurring conceptual difficulty is that several distinct ideas are often bundled together:
\begin{itemize}[leftmargin=2em]
\item \textbf{Kinematic nonseparability:} a state in $A\otimes B$ that does not factor as a product.
\item \textbf{Stochastic/operational realism:} in statistical optics one treats the field (or mode
amplitudes) as stochastic, and measurement outcomes as arising probabilistically from a
preparation+measurement procedure rather than as pre-assigned values.
\item \textbf{Contextuality:} the impossibility of a single joint assignment (or a single joint distribution)
whose marginals reproduce all observed context-wise statistics.
\end{itemize}

\begin{figure}[H]
  \centering
  \includegraphics[width=0.98\linewidth]{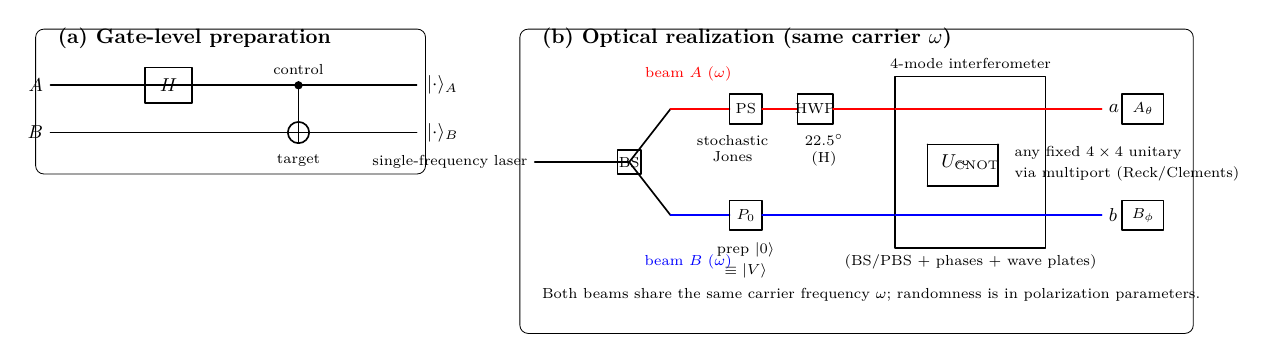}
  \caption{Bell-state preparation in two complementary views.
(a) Abstract circuit: a Hadamard on one input followed by a CNOT produces a Bell state.
(b) Optical realisation (schematic): a Hadamard-like splitting and a CNOT-like routing/flip,
implemented with standard linear-optical elements and subsequent recombination/conditioning,
prepares a Bell-like two-beam state.
In practice one uses coherent laser modes (for phase stability), while polarization stochasticity is
introduced by controlled modulation/scrambling of the Jones/Stokes variables; elements such as VP
are then understood as state-preparation (projection/conditioning) steps rather than unitary gates.}
  \label{fig:bell_prep}
\end{figure}

This paper focuses on classical \emph{stochastic} optics, where the operational reading ``no value until
measurement'' is natural, and where Bell--CHSH tests \cite{CHSH1969} can be phrased as \emph{noncontextuality}
tests rather than as locality tests. The central emphasis is that the relevant Bell-like state is not
introduced by algebraic fiat as a convenient tensor expression. It is \emph{prepared} by an explicit
physical arrangement which, in an idealized description, implements the logic of a familiar circuit:
a Hadamard-like transformation on one beam, followed by a CNOT-like coupling between the two
beams. Crucially, the experimental arrangement (routing, recombination, and/or conditioning on
specific output events) is what removes the terms that would appear in a naive expansion of an
arbitrary tensor product (Fig.~\ref{fig:bell_prep}).

A more optics-oriented presentation of the same basic preparation idea, together with a sketch of how
a CHSH--Bell inequality can be obtained in classical polarization optics from a noncontextuality axiom,
appears in the author's earlier preprint~\cite{GhoseIntersystem2024}. The present manuscript is self-contained and reorganizes the analysis in categorical terms, while
making explicit the \emph{indistinguishability} requirement (e.g.~a common carrier frequency) needed for a truly
Bell-like state in polarization to be operationally meaningful.

\medskip
\noindent\textbf{Experimental and modelling clarification:}
To realise the Hadamard--CNOT preparation as a \emph{coherent} (unitary) transformation on the encoded
two-level degree of freedom, the optical modes should be phase-stable coherent beams in practice
(e.g.\ narrowband laser light). We take the two parties to be \emph{two distinct spatial modes} (beams) $A$ and $B$, derived from a common \emph{single-frequency} carrier at $\omega$, so that auxiliary labels (in particular frequency) do not provide ``which-term'' information that could trivially defeat polarization nonseparability. The \emph{stochasticity}
is engineered only in the polarization parameters (Jones/Stokes variables)---for instance by driving an
electro--optic modulator (Pockels cell) or a polarization scrambler with a noise sequence. Equivalently,
one may regard each run (or each time window) as preparing a coherent polarization state with a
\emph{random Jones vector},
\[
\ket{\psi(t)} \;=\; \alpha(t)\,\ket{H}+\beta(t)\,\ket{V},
\]
where $\alpha(t)$ and $\beta(t)$ are stochastic processes but are phase-coherent within each realization.
The PBS/beam-splitter network then implements the Hadamard/CNOT logic as a linear (mode)
transformation on the encoded two-level degree of freedom \emph{for each realization}; the observed
statistics correspond to the resulting ensemble (generically a mixed state).

In this sense the sources need not be ``unpolarized'' in the strict statistical-optics sense:
a completely depolarized (incoherent) input would be described by a mixture and a PBS would merely
\emph{route} orthogonal components rather than implement a Hadamard-like \emph{unitary} on a qubit.
In the present stochastic-optics viewpoint, the preparation network acts linearly on each realization
of the (random) Jones vector; the resulting object is an \emph{ensemble} of coherent output states: for each realization the transformation is unitary, but averaging over the stochastic polarization parameters yields a (generically) mixed density operator whose Bell/CHSH signatures \cite{CHSH1969} depend on preserving the required coherence through the preparation.

\paragraph{Operational classicality (``classical'' in the present sense):}
Throughout, ``classical'' is used in an explicitly \emph{operational} sense that is compatible with
Hilbert-space polarization kinematics. Concretely, we assume the experiment lies within a subtheory
admitting a classical random-field / phase-space description, specified by:

\begin{enumerate}[label=(C\arabic*),leftmargin=2.2em]
\item \textbf{States: phase-space classical:}
The optical sources are restricted to states that admit a classical phase-space representation (e.g.\
coherent/thermal or appropriate mixtures), so that no empirically accessible nonclassical signatures
such as Wigner-negativity are invoked.\footnote{If one wishes a stronger classicality notion, one may
impose a nonnegative Glauber--Sudarshan $P$-representation (mixtures of coherent states).}

\item \textbf{Transformations: linear optics + controlled classical noise:}
All gates are implemented by linear optical elements (mode mixing, polarization rotations, attenuators)
possibly supplemented by externally imposed classical randomness (e.g.\ stochastic modulation of the
polarization/Jones parameters via a scrambler or EOM noise drive). No intrinsically quantum resources
(single-photon nonlinearity, non-Gaussian ancillae, etc.) are assumed.

\item \textbf{Measurements: classical interface:}
Readout is through an explicitly classical interface (intensity/click statistics modelled as coarse-grained
events). Outcomes are classical data: they can be copied, stored, and broadcast without invoking a
measurement-disturbance postulate.

\item \textbf{Contextuality as a consistency condition, not ``quantumness'':}
Given (C1)--(C3), Bell/CHSH is interpreted as a constraint on global consistency of context-wise
statistics (existence of a global section / joint distribution). Hence contextuality is a structural property
of the induced empirical model, not a synonym for microscopic quantumness.
\end{enumerate}

Category theory provides two complementary frameworks which, taken together, make the
situation transparent:
\begin{enumerate}[leftmargin=2em]
\item A \textbf{process-theoretic} (monoidal) framework for preparations and measurements, in which
the entire arrangement is a morphism in an operational category. A standard choice is Selinger's
CPM construction $\mathbf{CPM}(\mathbf{FHilb})$ \cite{Selinger2007}.
Nothing in the CPM construction presupposes quantization: it is a purely categorical way to represent
probabilistic mixing and conditioning on finite-dimensional complex Hilbert spaces, which in our setting are simply
Jones (polarization) spaces.

\item A \textbf{sheaf-theoretic} framework for contextuality \cite{AbramskyBrandenburger2011,CoeckeKissinger2017,CoeckePaquettePavlovic2009}, in which
the experiment induces an empirical model, and contextuality is precisely a gluing obstruction.
\end{enumerate}
We now develop both layers and connect them via a direct ``process $\Rightarrow$ empirical model''
bridge.

\section{Polarization systems and an entangling preparation}

\subsection{Two-beam polarization systems and a fixed tensor order}

Let $A \cong \mathbb{C}^2$ denote the polarization system of beam $a$ and $B \cong \mathbb{C}^2$ that of
beam $b$. Fix orthonormal bases $\{\ket{0}_A,\ket{1}_A\}$ and $\{\ket{0}_B,\ket{1}_B\}$ (e.g.\ vertical/horizontal).
We write the composite system as $A\otimes B$ and \emph{fix this order once and for all}.

\subsection{Idealized target: Hadamard then CNOT}

The adjective ``Hadamard-like'' refers to a coherent $2\times 2$ basis transformation on the
\emph{encoded} two-level degree of freedom (e.g.\ a half-wave plate at $22.5^\circ$ acting on polarization,
or a balanced interferometer acting on a path/mode qubit). In the envisaged stochastic-optics regime, coherence is provided by stable laser modes, while randomness is engineered in the polarization parameters, so that the desired linear transformation is implemented pointwise on each realisation.

Define the Hadamard operator $H$ on $A$ by
\[
H\ket{0}_A=\frac{1}{\sqrt2}\big(\ket{0}_A+\ket{1}_A\big),\qquad
H\ket{1}_A=\frac{1}{\sqrt2}\big(\ket{0}_A-\ket{1}_A\big),
\]
and the CNOT on $A\otimes B$ (control $A$, target $B$) by
\[
\mathrm{CNOT}\,\ket{x}_A\ket{y}_B=\ket{x}_A\ket{y\oplus x}_B.
\]
Then
\[
\ket{0}_A\ket{0}_B
\ \xmapsto{\,H\otimes \id_B\,}\
\frac{1}{\sqrt2}\big(\ket{0}_A+\ket{1}_A\big)\ket{0}_B
\ \xmapsto{\,\mathrm{CNOT}\,}\
\ket{\Phi^+}\;:=\;\frac{1}{\sqrt2}\big(\ket{0}_A\ket{0}_B+\ket{1}_A\ket{1}_B\big).
\]
In optical realisations, the Hadamard and CNOT are implemented by concrete linear optical elements
and path/polarization manipulations; the point here is that the \emph{entangling gate} is the physical
mechanism that produces nonfactorisability at the output.

\begin{remark}[Physical ``term elimination'' is part of the preparation]
A generic tensor product of two generic single-beam polarization kets expands into four terms.
The two-term Bell-like form is not obtained by algebraic deletion; it is obtained because the
experimental arrangement implements an entangling coupling and (explicitly or implicitly)
conditions on an output event (port selection, coincidence class, etc.). We formalize this as
conditioning on a flag/ancilla system in the process layer.
\end{remark}


\subsection{Alternative preparation via external conical refraction (ECR)}
As an experimentally distinct route to the same two-beam ``Bell-like'' sector, one may exploit external conical
refraction in a biaxial crystal. A collimated input beam sent along an optic axis emerges as a bright ring; crucially,
each pair of diametrically opposite points on the ring carries orthogonal linear polarizations \cite{BerryJeffreyLunney2006}.
Moreover, cascaded conical refraction through multiple crystals can generate multiple concentric rings \cite{Turpin2013}, offering
additional geometric degrees of freedom for selection and interference.

By selecting and interfering suitable sectors of (one or two) such rings, and applying the same kind of routing/conditioning
used in the Hadamard--CNOT scheme, one can prepare a high-visibility two-beam polarization coherence matrix that approximates
a Bell state in the retained branch. (For related cascade implementations in applied settings see, e.g., \cite{ODwyer2012}.)
Conceptually, the geometry provides a classical analogue of the intersecting emission cones in spontaneous parametric down conversion (SPDC),
with selected ECR ring sectors playing the role of ``cone intersections.''

\begin{figure}[H]
  \centering
  \includegraphics[width=0.95\linewidth]{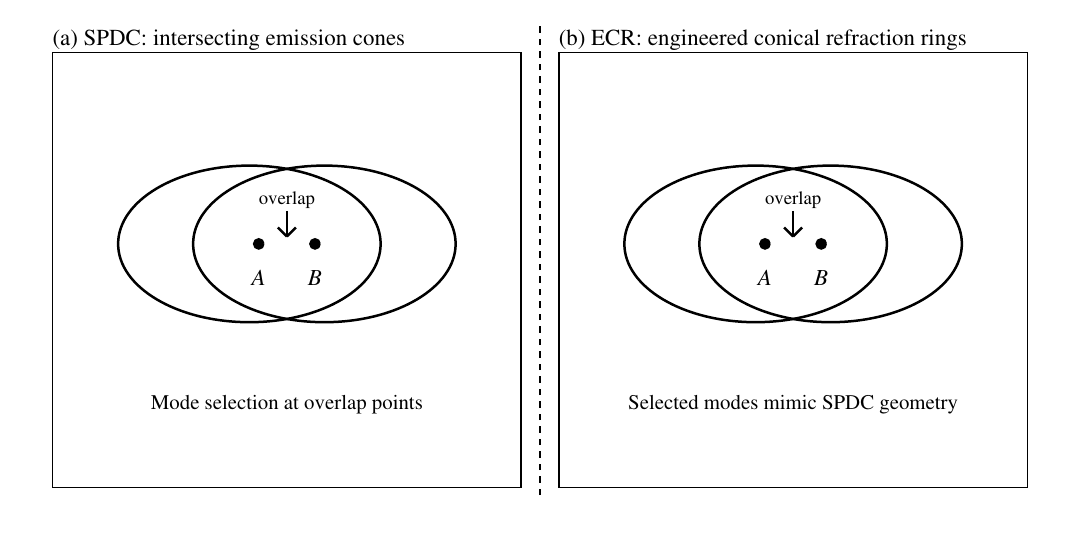}
  \caption{Schematic comparison: intersecting SPDC emission cones (left) and an ECR-based analogue (right), where intersecting
  conical-refraction ring sectors and subsequent routing/conditioning select a Bell-like two-beam polarization branch.}
  \label{fig:ecr_spdc_mimic}
\end{figure}

This alternative does not change the logical structure of the later analysis: the preparation is still best modelled as a flagged
CP instrument, and the CHSH/sheaf-theoretic discussion applies to the conditioned subensemble it prepares.

\section{Process layer: $\mathbf{CPM}(\mathbf{FHilb})$ and conditioning}

\subsection{Ambient category and operational primitives}

We work in an operational symmetric monoidal category of processes. A concrete choice is
$\mathbf{CPM}(\mathbf{FHilb})$ \cite{Selinger2007}:
objects are finite-dimensional Hilbert spaces (system types), morphisms are completely positive maps,
$\otimes$ composes systems in parallel, and discarding corresponds to the trace.

A (possibly mixed) state of $X$ is a morphism $I\to X$, i.e.\ a density operator $\rho_X$.
An \emph{effect} is a morphism $X\to I$ (a yes/no event or scalar test). A measurement is an
\emph{instrument} (family of CP maps) with total map trace-preserving.

\subsection{Preparation as a single morphism with a flag system}

Let $\rho_{\mathrm{in}}$ be the (possibly stochastic) input state on $A\otimes B$ before the network.
Introduce an auxiliary system $E$ which records the output event class produced by routing/detection.
The complete preparation is represented as a CP map
\[
\mathcal{U}:\ \rho_{\mathrm{in}} \longmapsto \rho_{ABE}.
\]

\begin{definition}[Conditioned (postselected) output]
Let $e:E\to I$ be the effect representing the event retained by the experimental protocol.
The conditioned two-beam state is
\[
\rho^{(e)}_{AB}
\ :=\
\frac{(\id_{AB}\otimes e)\big(\rho_{ABE}\big)}
{\Tr\!\big[(\id_{AB}\otimes e)\big(\rho_{ABE}\big)\big]}.
\]
\end{definition}

\paragraph{Interpretation of the flag system $E$:}
The auxiliary system $E$ is not a hidden variable; it is a \emph{classical record} produced by the
preparation-and-routing hardware (output port, trigger window, accepted vs.\ rejected runs, etc.).
Equivalently, the preparation is an \emph{instrument} $\{\mathcal{E}_e\}_e$ whose branches occur with probabilities
$p(e)=\Tr\,\mathcal{E}_e(\rho_{\mathrm{in}})$ and prepare corresponding subensemble states
$\rho^{(e)}_{AB}=\mathcal{E}_e(\rho_{\mathrm{in}})/p(e)$. Writing a discrete label $e$ does not imply individual
microscopic ``events''; in stochastic optics the probabilities arise from ensemble statistics of the fluctuating classical field
(and detector noise) under the applied conditioning rule.

In the idealized pure-state limit, $\rho^{(e)}_{AB}\approx \proj{\Phi^+}$ (up to basis conventions and a global phase).
In real experiments, imperfections and partial coherence yield a mixed state; the CPM framework is designed to
accommodate this without changing the logic of the argument.

\subsection{Local measurement settings and correlations}

For two-outcome polarization measurements at angle $\theta$ on $A$ and $\phi$ on $B$, define
\[
\ket{+_\theta}=\cos\theta\,\ket{0}+\sin\theta\,\ket{1},
\qquad
\ket{-_\theta}=-\sin\theta\,\ket{0}+\cos\theta\,\ket{1},
\]
with projectors $\Pi_{\pm|\theta}:=\proj{\pm_\theta}$. In classical optics the corresponding detected
intensities satisfy Malus' law: for a beam prepared with linear polarization at angle $\theta_0$,
the (normalized) transmitted intensity is $I_{+}(\theta)=\cos^2(\theta-\theta_0)\in[0,1]$ and
$I_{-}(\theta)=\sin^2(\theta-\theta_0)\in[0,1]$. For a general (possibly stochastic) preparation,
one may work with normalized intensities (probabilities) obtained from the reduced states
$\rho^{(e)}_{A}:=\Tr_B\rho^{(e)}_{AB}$ and $\rho^{(e)}_{B}:=\Tr_A\rho^{(e)}_{AB}$:
one may work with normalized intensities (probabilities)
\[
p(\pm|\theta)=\Tr\!\big(\rho^{(e)}_{A}\,\Pi_{\pm|\theta}\big),\qquad p(+|\theta)+p(-|\theta)=1,
\]
and define the \emph{contrast} (a bounded real outcome)
\[
A_\theta \;:=\; \frac{I_{+}(\theta)-I_{-}(\theta)}{I_{+}(\theta)+I_{-}(\theta)}
\;=\; p(+|\theta)-p(-|\theta)\ \in[-1,1].
\]
Equivalently, introducing the associated dichotomic observable
\[
\sigma_\theta := \Pi_{+|\theta}-\Pi_{-|\theta}
= \cos 2\theta\,\sigma_z+\sin 2\theta\,\sigma_x,
\quad\text{where }\sigma_z=\proj{0}-\proj{1}\text{ and }\sigma_x=\ket{0}\!\bra{1}+\ket{1}\!\bra{0}.
\]
one has $\langle A_\theta\rangle=\Tr(\rho^{(e)}_A\,\sigma_\theta)$. Defining $B_\phi$ analogously on $B$,
the correlation function can be written either as the average product of contrasts,
\[
E(\theta,\phi)\;:=\;\langle A_\theta\,B_\phi\rangle \in[-1,1],
\]
or, in operator form, as
\[
E(\theta,\phi)\;=\;\Tr\!\big(\rho^{(e)}_{AB}\,(\sigma_\theta\otimes \sigma_\phi)\big).
\]
This is the operational quantity extracted from intensities/coincidence statistics in the stochastic-optics setting.

\section{Sheaf layer: contextuality as failure of gluing}

\subsection*{Measurement scenario, presheaf, and empirical model}

A measurement scenario consists of:
\begin{itemize}[leftmargin=2em]
\item a finite set $X$ of measurement labels (settings),
\item a family $\mathcal{M}$ of contexts (jointly implementable sets of measurements),
\item an outcome set $O$. In many textbook CHSH presentations one takes $O=\{\pm1\}$,
but in classical polarization optics the primitive observables are intensities. By Malus' law,
the transmitted intensity through a polarizer at angle $\theta$ is proportional to
$\cos^2(\theta-\theta_0)$; after normalization one has $I(\theta)\in[0,1]$. It is therefore natural
to allow real-valued outcomes, and to work either directly with $O\subseteq[0,1]$ (intensities)
or with normalized contrasts taking values in $[-1,1]$.
\end{itemize}

\noindent
For CHSH-type inequalities the essential assumption is boundedness: if each local quantity used
to form correlations is bounded in absolute value by $1$ (as the Malus-law normalization ensures),
then any noncontextual (global-section) model satisfies $|S|\le 2$.

For CHSH one may take $X=\{a,a',b,b'\}$ and
\[
\mathcal{M}=\big\{\{a,b\},\{a,b'\},\{a',b\},\{a',b'\}\big\}.
\]
Define the event presheaf $\mathcal{E}$ by $\mathcal{E}(C)=O^C$ for each $C\in\mathcal{M}$ with restriction maps
given by forgetting components (equivalently, marginalizing assignments).

\begin{definition}[Empirical model \cite{AbramskyBrandenburger2011}]
An empirical model is a family $\{p_C\}_{C\in\mathcal{M}}$, where $p_C$ is a probability distribution on
$\mathcal{E}(C)$, and the family is compatible under marginalization on overlaps.
\end{definition}

\begin{definition}[Noncontextuality as a global section \cite{AbramskyBrandenburger2011}]
The empirical model is \emph{noncontextual} if there exists a global distribution $p_X$ on $O^X$ whose marginals
reproduce $p_C$ for every $C\in\mathcal{M}$. Equivalently, the model admits a global section (a consistent gluing).
\end{definition}

The CHSH inequality is a remarkably robust constraint. Although it is often introduced via
local hidden-variable models, the bound $|S|\le 2$ follows under a variety of assumption sets.
In particular, it holds whenever one has (i) a single global joint distribution reproducing the
context-wise statistics (equivalently, a global section / noncontextual model), and (ii) bounded
local quantities $|A_\theta|\le 1$, $|B_\phi|\le 1$ (as ensured here by Malus-law normalization).
Thus, in the present setting CHSH is most naturally read as a constraint of global consistency
(noncontextuality) rather than as a specifically ``nonlocality'' postulate.

\subsection*{Bridge: processes induce empirical models}

Given $\rho^{(e)}_{AB}$ and a specification of measurement instruments for each setting, the process theory computes
the context-wise distributions $p_C$ by composition. Thus:
\[
\text{(CPM state + instruments)} \quad\Rightarrow\quad \text{empirical model (presheaf data)}.
\]
Contextuality is then assessed at the level of the induced empirical model.

\section{CHSH: product states glue; Bell-like outputs can fail to glue}

\subsection*{CHSH bound from global sections}

Define the CHSH expression
\[
S\;:=\;E(\theta,\phi)+E(\theta,\phi')+E(\theta',\phi)-E(\theta',\phi').
\]

\begin{lemma}[CHSH bound from global sections]
If the empirical model admits a global section (equivalently, a joint distribution on $O^X$),
then $|S|\le 2$.
\end{lemma}

This is the familiar CHSH constraint \cite{CHSH1969}. In the sheaf formulation it is a direct consequence of the
existence of a single joint distribution reproducing all context-wise distributions \cite{AbramskyBrandenburger2011}.

\subsection*{Product states are noncontextual}

If $\rho^{(e)}_{AB}=\rho_A\otimes\rho_B$ and the measurements are local, then the induced correlations are compatible
with a joint distribution; the empirical model admits a global section, hence $|S|\le 2$.

\begin{proposition}[Product states glue]
For local two-setting/two-outcome measurements, any empirical model induced by a product state $\rho_A\otimes\rho_B$
is noncontextual and satisfies $|S|\le 2$.
\end{proposition}

\subsection*{Bell-like conditioned outputs and maximal violation}

For the ideal Bell-like state $\ket{\Phi^+}=\frac{1}{\sqrt2}(\ket{00}+\ket{11})$, one finds (with the above conventions)
\[
E(\theta,\phi)=\cos\big(2(\theta-\phi)\big),
\]
so that choosing
\[
\theta=0,\quad \theta'=\frac{\pi}{4},\quad
\phi=\frac{\pi}{8},\quad \phi'=-\frac{\pi}{8}
\]
yields
\[
S=2\sqrt2.
\]
Hence the induced empirical model violates CHSH and therefore has no global section.

\begin{proposition}[CHSH violation $\Rightarrow$ no global section]
For the Bell-like conditioned output state (ideal limit) and suitable measurement choices, the induced empirical model
violates CHSH and therefore admits no global section. Hence it is contextual in the sheaf-theoretic sense.
\end{proposition}

In this two-level view, ``Bell-like nonseparability'' is a property of the \emph{prepared process} producing
$\rho^{(e)}_{AB}$ in a non-cartesian monoidal setting, while ``CHSH violation'' is a property of the \emph{empirical model}
extracted from measurement statistics. The latter is captured precisely as a failure of gluing (no global section),
i.e.\ contextuality.

\section{Discussion and conclusion}

\subsection*{A shift in what ``classical reality'' means}

In the present framework, contextuality is defined operationally as a failure of a global section of the
event presheaf, i.e.\ the nonexistence of a single joint distribution whose marginals reproduce all
context-wise statistics. What is conceptually striking here is that the underlying optical resources may
remain ``classical'' in the phase-space/measurement sense adopted throughout (no empirically accessible
nonclassical signatures are assumed), yet the induced empirical model can still be \emph{contextual}.
Thus the identification ``classical $\Rightarrow$ noncontextual'' is not a logical necessity; rather,
noncontextuality is an additional assumption about context-independent reality. In this sense the work
motivates a refined notion of classical reality: one may have a classical operational substrate while
global, context-independent value assignments fail once incompatible measurement contexts are taken
seriously.

\subsection*{Scope and limitations}
The point of the stochastic-optics Bell preparation is not to claim ``quantumness by fiat,'' but to
separate, in a controlled and fully operational way, three layers that are often conflated:
(i) Hilbert-space \emph{kinematics} (tensor-product nonfactorisability of prepared states),
(ii) the \emph{process} description of preparation/conditioning/measurement, and
(iii) the \emph{logic of consistency across contexts} (global sections versus failure-to-glue).
Because the preparation network is explicit and tunable, it provides a concrete testbed for probing
how Bell--CHSH witnesses behave under noise, coarse-graining, imperfect coherence, and
postselection---and for clarifying what exactly is (and is not) certified by a CHSH violation in an
operational, stochastic setting.

\subsection*{What it cannot do (without extra physical assumptions)}

The present framework is deliberately clinical about limitations. First, the copyable/deletable
(classical) interface applies to measurement \emph{records} (intensities/clicks), not to the prepared
Bell-like state itself; nevertheless, the availability of a classical interface typically prevents one from
importing unconditional quantum-cryptographic guarantees \emph{solely} from Bell--CHSH statistics.
Second, a CHSH violation here witnesses a failure of noncontextual global assignment for the induced
empirical model; it does not, by itself, entail the full suite of quantum-information constraints (such as
monogamy-based device-independent secrecy) unless additional assumptions are imposed on the
adversary's access to the underlying field/process. Finally, the analysis does not claim automatic
quantum computational speedups: it identifies a structural and operational analogue of Bell resources,
and clarifies which further principles would be needed to promote it to specifically quantum
information-processing power.

\subsection*{Category theory helps separate what is often conflated}

\begin{itemize}[leftmargin=2em]
\item \textbf{Kinematic nonseparability} is a monoidal notion: in Hilbert-style tensor products, nonfactorisable states
exist independently of any measurement postulate.
\item \textbf{Operational stochasticity} is naturally expressed in a CP/process setting: states are probabilistic objects,
and the preparation/measurement architecture (including conditioning on events) is part of the morphism.
\item \textbf{Contextuality} is a property of the induced empirical model: CHSH violation is a precise obstruction to global
sections.
\end{itemize}

This clarifies why, in classical stochastic optics, the physically correct comparison is not between ``a naive tensor
product of two independent beams'' and a Bell ket, but between the \emph{conditioned output} $\rho^{(e)}_{AB}$ of an
entangling preparation (Hadamard-like splitting + CNOT-like coupling + routing/conditioning) and a Bell-like state.
Once this is made explicit, it is unsurprising that one can obtain CHSH violations of quantum strength at the level of
correlations.

\appendix
\section{Appendix: string-diagram view of preparation, conditioning, and statistics}

This appendix gives a compact diagrammatic rendering (in the spirit of monoidal ``process theories'')
of the two key steps emphasized in the main text:
(i) physical elimination of unwanted terms via a flag/ancilla and conditioning, and
(ii) extraction of an empirical model from a state and local measurements.

\subsection{Preparation with a flag and conditioning}

We depict a preparation as a box $\mathcal U$ that takes an input state on $AB$ and produces an
output on $AB$ together with a classical flag $E$ that records which event class occurred. Conditioning
is an effect $e$ applied to $E$.

\begin{center}
\begin{tikzpicture}[x=1.1cm,y=0.9cm,>=Latex]
  \draw (-3,1) -- (-1,1) node[midway,above] {$A$};
  \draw (-3,0) -- (-1,0) node[midway,below] {$B$};

  \draw (-1.2,-0.4) rectangle (0.2,1.4);
  \node at (-0.5,0.5) {$\mathcal U$};

  \draw (0.2,1) -- (2.4,1) node[midway,above] {$A$};
  \draw (0.2,0) -- (2.4,0) node[midway,below] {$B$};

  \draw (0.2,2) -- (1.4,2) node[midway,above] {$E$};

  \draw (1.4,1.7) rectangle (2.1,2.3);
  \node at (1.75,2.0) {$e$};

  \draw (2.1,2) -- (2.9,2);
  \node at (3.15,2.0) {$I$};

  \node at (-3.0,0.5) {$\rho_{\mathrm{in}}$};
  \node at (2.35,0.5) {$\rho_{AB}^{(e)}$};
\end{tikzpicture}
\end{center}

Algebraically this corresponds to
\[
\rho^{(e)}_{AB}
\;\propto\;
(\id_{AB}\otimes e)\big(\rho_{ABE}\big),
\qquad
\rho_{ABE}=\mathcal U(\rho_{\mathrm{in}}).
\]
The normalization constant is the success probability of the event $e$.

\subsection{From a conditioned state to context-wise probabilities}

Let $x$ be a setting on $A$ and $y$ a setting on $B$. A two-outcome measurement is represented by
effects $\{m^{A}_{o|x}\}_{o\in\{\pm\}}$ on $A$ and $\{m^{B}_{o'|y}\}_{o'\in\{\pm\}}$ on $B$.
The joint probability is
\[
p(o,o'\,|\,x,y)=\Tr\!\big(\rho_{AB}^{(e)}\,(m^{A}_{o|x}\otimes m^{B}_{o'|y})\big).
\]

\begin{center}
\begin{tikzpicture}[x=1.1cm,y=0.9cm,>=Latex]
  \draw (-2.6,-0.4) rectangle (-1.2,1.4);
  \node at (-1.9,0.5) {$\rho_{AB}^{(e)}$};

  \draw (-1.2,1) -- (0.0,1) node[midway,above] {$A$};
  \draw (-1.2,0) -- (0.0,0) node[midway,below] {$B$};

  \draw (0.0,0.7) rectangle (0.9,1.3);
  \node at (0.45,1.0) {$m^{A}_{o|x}$};

  \draw (0.0,-0.3) rectangle (0.9,0.3);
  \node at (0.45,0.0) {$m^{B}_{o'|y}$};

  \draw (0.9,1) -- (1.6,1);
  \draw (0.9,0) -- (1.6,0);
  \node at (1.85,0.5) {$p(o,o'|x,y)$};
  \draw (1.6,1) -- (1.6,0);
\end{tikzpicture}
\end{center}

Varying $(x,y)$ over the CHSH contexts yields the family of distributions $\{p_C\}_{C\in\mathcal M}$,
i.e.\ the empirical model to which the sheaf-theoretic global-section criterion applies.

\section{Acknowledgements}
I acknowledge helpful correspondence with C. S. Unnikrishnan regarding Ref. \cite{GhoseIntersystem2024} and
use of AI tools for language polishing and structing. However, I am solely rsponsible for the content.

\end{document}